\documentclass[runningheads]{llncs}
\usepackage{amsmath}
\usepackage{graphicx}
\usepackage{booktabs}
\usepackage{hyperref}
\usepackage[T1]{fontenc}
\usepackage{microtype}
\usepackage{float}
\usepackage[shortlabels]{enumitem}
\usepackage{orcidlink}
\hypersetup{hidelinks}
\begin{document}
\raggedbottom
\title{Harnessing Disagreement: Detecting Correlated Agreement Blindness in Multi-Agent Triage\thanks{Accepted for publication at
PAAMS 2026. This is the pre-review submitted version.}}

\titlerunning{Harnessing Disagreement in Multi-Agent Triage}

\author{Shay Seiya McDonnell\inst{1,2}\orcidlink{0009-0007-4084-7897} \and
Avantika Singh\inst{3} \and
Quoc-Viet Pham\inst{1,2}\orcidlink{0000-0002-9485-9216} \and
Vratislav Havl\'{i}k\inst{3} \and
Gregory M.P. O'Hare\inst{1,2}\orcidlink{0000-0002-5124-1686}}

\authorrunning{S.\,S. McDonnell et al.}

\institute{
School of Computer Science \& Statistics, Trinity College Dublin, College Green, Dublin 2, Ireland\\
\email{mcdonns8@tcd.ie, viet.pham@tcd.ie, gregory.ohare@tcd.ie}
\and
ADAPT Centre, Trinity College Dublin, College Green, Dublin 2, Ireland
\and CKDelta, 28/29 Sir John Rogerson's Quay, Dublin 2, Ireland\\
\email{Avantika.Singh@ckdelta.ai, Vratislav.Havlik@ckdelta.ai}
}

\maketitle
\begin{abstract}
Disagreement-triggered escalation can create a structural blind spot in multi-agent arbitration: as base learners improve, they tend to converge, weakening safety monitoring where correlated failures concentrate. We term this \textit{correlated agreement blindness} and present ARAT (Arbitrated Reasoning Agents for Alarm Triage), a directed-star system combining an inductive Random Forest (RF) agent, an analogical case-based k-nearest neighbour (k-NN) agent, and a calibrated meta-model to mitigate this effect. On 82,332 holdout samples from the UNSW-NB15 network intrusion detection dataset, 57.2\% of errors occur under agreement and 90.6\% of dangerous under-predictions evade disagreement-based monitoring even after conservative override; ablation shows that strengthening base learners increases error correlation while reducing disagreement. ARAT reduces under-prediction relative to soft voting from 4.80\% to 1.70\% via conservative override ($-$2.6pp) and a safety-flag gate ($-$0.5pp), demonstrating architectural gains. Cross-dataset validation on clinical readmission supports these indicators, suggesting that diversification improves safety only when it generates productive disagreement rather than convergence. These results indicate that disagreement-triggered escalation can be blind to correlated failure, a risk that may intensify as agentic pipelines deploy increasingly capable, correlated models.

\keywords{Multi-agent systems \and Correlated agreement blindness \and
Disagreement uncertainty \and Network intrusion detection \and
Agentic AI \and Human-in-the-loop}

\end{abstract}

\section{Introduction}
\label{sec:intro}

Disagreement is widely used as a practical proxy for uncertainty in
multi-agent systems (MAS)~\cite{seung1992qbc,wooldridge2009}. As
agentic AI pipelines deploy increasingly capable base models, those
models agree more often, and that agreement becomes a misleading
safety signal when the models share correlated errors~\cite{kim2025correlated,kuai2026independent}. Unlike prior
work on error correlation in ensemble
accuracy~\cite{brown2005diversity,wood2023diversity}, we show this
blindness is heightened by accuracy improvements: stronger
models suppress the very disagreement signal that triggers human review~\cite{goel2025great}. The resulting failure is categorical:
when agents mistakenly agree, escalation is never triggered and
dangerous cases bypass human review entirely.

This paper makes the following contributions:
\begin{enumerate}[leftmargin=*,nosep]

\item \textbf{Characterisation:} Large-scale empirical evidence of
\textit{correlated agreement blindness}: 3.53$\times$ joint error
inflation (bias-corrected and accelerated (BCa) 95\% confidence interval (CI): [3.50, 3.59]), error association $\phi=0.612$ (BCa 95\% CI:
[0.605, 0.619]) at $n{=}82{,}332$, and 90.6\% of dangerous under-predictions
not detected by disagreement-based monitoring. A paired ablation
demonstrates that $\phi$ increases and disagreement decreases as the RF is made heavier.

\item \textbf{Architecture:} ARAT, a directed-star arbitration system
with two quantifiable routing layers: conservative override
($-$2.6pp under-prediction vs.\ soft voting) and unanimous-Normal safety flag ($-$0.5pp further, exact: $-$0.48pp), plus a calibrated escalation meta-model (area under the receiver operating characteristic curve AUROC 0.924); together consistently outperforming evaluated single-model and
naive MAS baselines and not reproduced by loss-function tuning alone.

\item \textbf{Performance:} 98.3\% operationally adjusted accuracy (treating over-predictions as acceptable and penalising only under-predictions) and
1.70\% under-prediction on 82,332 UNSW-NB15 holdout samples, with a
per-layer decomposition, indicating that each architectural layer
contributes independently.

\item \textbf{Validation:} Cross-dataset replication on UCI Diabetes
($n=20{,}354$) supports the same mechanism indicators; SVM substitution
shows that agent diversification improves safety only when it generates
productive disagreement: on UNSW-NB15 SVM diverges and helps, whereas on
Diabetes it converges and does not.

\end{enumerate}

\section{Related Work}
\label{sec:related}

Prior ML-based network intrusion detection work~\cite{moustafa2015unsw,moustafa2016evaluation} rarely addresses uncertainty quantification for human-in-the-loop severity triage. Cost-sensitive approaches encode asymmetric costs directly~\cite{elkan2001cost,frank2001ordinal}; consensus formation in MAS has been extensively studied as a coordination objective~\cite{olfatiSaberFaxMurray2007}. In contrast ARAT applies conservative post-hoc arbitration, separating predictive and routing concerns. Error correlation is well established~\cite{brown2005diversity,wood2023diversity}, but prior work studies its effect on accuracy rather than escalation-triggered review. Goel et al.~\cite{goel2025great} demonstrate that capable models share similar blind spots, which undermines AI-based oversight. However, they examine an \textit{AI-as-evaluator} setting, where one model is tasked with grading another; in this case, the oversight is merely weakened. By contrast, our work focuses on \textit{disagreement-triggered routing}, where human review is gated by model consensus. Here, if agents mistakenly agree, the safety mechanism fails entirely and the human analyst is never alerted. Thus, our failure mode is categorical (a total loss of oversight) rather than quantitative (an inaccurate evaluation). Hammond et al.~\cite{hammond2025multiagent} identify correlated error suppression of divergence as a multi-agent risk; Reid et al.~\cite{reid2025risk} identify monoculture collapse as a critical vulnerability in governed MAS. Our contribution is orthogonal to both: we demonstrate that the blindness is not merely present but intensifies as base learner accuracy improves, and we quantify the architectural layers required to mitigate it. Concurrent work confirms the breadth of correlated failure: larger
LLMs exhibit highly correlated errors even across distinct architectures~\cite{kim2025correlated}, latent entanglement induces over-endorsement bias in verifier ensembles~\cite{kuai2026independent}, and correlated ensembles face an information-theoretic error floor that additional models cannot overcome~\cite{turkmen2026ensemble}. These studies address ensemble accuracy or evaluation bias, not escalation-triggered human review. To our knowledge, this failure mode has not been explicitly characterised in escalation-triggered multi-agent systems, nor has its relationship with increasing model accuracy been systematically examined.

\section{Use Case: Safety-Critical Triage}
\label{sec:usecase}

ARAT addresses triage settings where uncertainty signals prioritise
human review rather than trigger abstention. In Security Operations
Centres (SOCs), analysts triage alerts by severity: under-prediction
(predicting lower severity than truth) is dangerous; over-prediction
erodes trust. Under-prediction rate is therefore our primary safety
metric; the same structure applies wherever dangerous minority cases
must be prioritised under resource constraints. UNSW-NB15 was selected because its attack taxonomy closely mirrors operational alert categories and its scale and label structure make it suitable for evaluating severity triage in a safety-critical setting. This work is conducted in the context of an active industrial collaboration in which ARAT has been deployed on a similar task.

\section{ARAT Architecture}
\label{sec:architecture}

In its current incarnation, ARAT comprises two predictive agents
(Random Forest and $k$-NN) and a central routing agent that uses
disagreement and a learned risk score to arbitrate between predictions
and escalate uncertain cases for operator review. ARAT provides a
mechanistic and architectural contribution: the goal is to demonstrate
that correlated agreement blindness is a structural property of
disagreement-based escalation, and to show that a deliberately
lightweight routing design can mitigate it without retraining base
models. The routing agent $RA_3$ is intentionally simple: deterministic
rules for the first two safety layers and logistic regression for the
third, to maximise interpretability and auditability in analyst-facing
deployments where opaque arbitration is operationally unacceptable.
This follows the committee-style MAS
tradition~\cite{seung1992qbc,wooldridge2009}, prioritising principled
routing logic over negotiation or belief revision.

\begin{figure}[!htbp]
\vspace{4pt}
\centering
\includegraphics[width=1\linewidth]{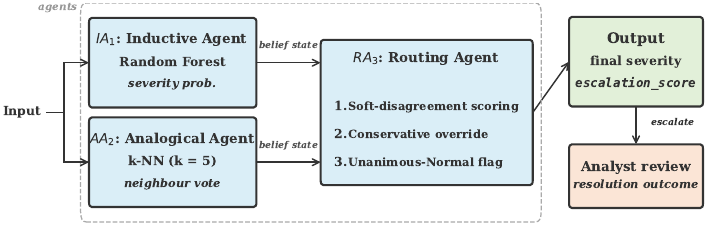}
\vspace{-10pt}
\caption{ARAT directed-star topology. $IA_1$ (RF) and $AA_2$ ($k$-NN)
produce belief states processed by $RA_3$ across three stages:
(1)~soft-disagreement scoring (weighted sum of Shannon entropy
and binary disagreement), (2)~conservative severity override on
disagreement, and (3)~unanimous-Normal safety flag when entropy
$> \theta=0.10$. $RA_3$ emits a severity prediction and calibrated
\texttt{escalation\_score}; flagged cases escalate to mandatory review.}
\label{fig:arch}
\end{figure}

ARAT implements a directed-star topology (Figure~\ref{fig:arch}) in which $IA_1$ and $AA_2$
report belief states to a central routing agent, $RA_3$.
The topology is agent-agnostic (any classifier emitting class probabilities
can join the star), and while demonstrated on a minimal two-agent
configuration, the blindness mechanism generalises to any topology where
a subset of agents can reach consensus on a wrong prediction undetected.

\paragraph{$IA_1$: Inductive agent (Random Forest).}
$IA_1$ is a Random Forest classifier trained on the pre-holdout
training split using $n=30$ mutual-information (MI) selected features
from the UNSW-NB15 dataset. It uses 500 trees with no depth
restriction, minimum leaf size of 2, and balanced class weights to
account for the 0.51\% prevalence of the High-severity class. All
baselines use the identical feature set. $IA_1$ outputs
class-probability vectors from tree vote fractions as confidence
signals.

\paragraph{$AA_2$: Analogical agent ($k$-NN).}
$AA_2$ implements analogical case-based retrieval via $k$-nearest
neighbour ($k=5$, uniform weighting) using the same $n=30$
MI-selected features. Neighbours are retrieved by Euclidean distance;
probability vectors for disagreement and entropy routing are
constructed from unweighted neighbour-label frequencies. $AA_2$
operates independently of $IA_1$. Despite this epistemic distinction,
both agents share the same feature space and training distribution,
which underlie the high error correlation observed in Section~\ref{sec:dependence}:
independence of method does not guarantee independence of failure. While brute-force $k$-NN inference over $n=175{,}341$ training points is computationally expensive, approximate nearest-neighbour methods (e.g., FAISS~\cite{johnson2017billion}) provide a drop-in, low-latency replacement that preserves our routing logic and architectural properties.

\paragraph{$RA_3$: Routing agent.}
$RA_3$ arbitrates using a soft-disagreement score $c = 0.5 \cdot H(p_{\text{merged}})+
0.5 \cdot d$, where $H(p_{\text{merged}})$ is Shannon entropy over the averaged class-probability
distributions and $d=1$ if agents predict different classes ($d=0$ otherwise). Both components are
min-max scaled to $[0,1]$. Equal weighting is the stable default; ablation over
asymmetric alternatives (0.3/0.7, 0.7/0.3) showed no consistent improvement ($\leq$0.1pp difference).

On disagreement, $RA_3$ applies a conservative severity override,
choosing the more severe prediction to prioritise safety and eliminate 92.8\% of
disagreement-based under-predictions. When both agents unanimously predict the lowest-risk class and entropy
$>\theta=0.10$, a safety flag triggers mandatory analyst review;
$\theta=0.10$ was selected to flag approximately 6.79\% of cases;
this threshold is configurable and was chosen to reflect a plausible
analyst review budget in operational SOC deployments. The escalation meta-model (logistic regression,
10 routing-time features, isotonic calibration) provides a calibrated
$P(\text{under-prediction})$ score for analyst queue ordering
(AUROC 0.924).

\section{Experimental Setup}
\label{sec:experiments}

\paragraph{Dataset and split.}
We evaluate on the UNSW-NB15 network intrusion detection dataset from
UNSW Canberra's Cyber Range
Lab~\cite{moustafa2015unsw,moustafa2016evaluation,moustafa2017geometric,moustafa2017dirichlet,sarhan2020netflow},
comprising 2,540,044 total network traffic records combining real
normal activities with synthetically generated attacks. We use the
official partition: training set ($n=175{,}341$,
\texttt{UNSW\_NB15\_training-set.csv}) and testing set ($n=82{,}332$,
\texttt{UNSW\_NB15\_testing-set.csv}), mapping native labels into our
4-class severity triage scheme (Normal, Low, Medium, High). The
High-severity class represents 0.51\% of our holdout set (422 cases).

\paragraph{Cross-dataset validation.}
To evaluate the generalisability of correlated agreement blindness beyond intrusion detection, we replicate ARAT on the UCI Diabetes 130-US Hospitals dataset~\cite{strack2014diabetes}. We apply a fixed
80/20 split (train $n=81{,}412$, test $n=20{,}354$) and map
readmission outcomes into a three-class severity scheme (test set: No
readmission 60.4\%, Readmitted $>$30 days 28.9\%, and Readmitted
$<$30 days 10.6\%), with the last class as the dangerous
under-prediction target. Twenty-one features are
available at discharge time; none are post-hoc. The Diabetes pipeline
uses adapted base learner configurations ($n=100 \rightarrow 500$
trees, fixed depth $=15$, \texttt{min\_samples\_leaf}${}={}$5) for this noisier task. The routing architecture,
conservative override, and safety flag use the same logic as
UNSW-NB15.

\paragraph{Features.}
For UNSW-NB15 we use $n=30$ MI-selected tabular
features~\cite{peng2005mi}, removing noisy features identified via
a feature-count sweep optimising High-class F1. For Diabetes we use
all 21 discharge-time features after excluding identifiers.

\paragraph{Metrics.}
We report exact accuracy as a conventional summary metric, but our primary evaluation is safety-oriented: over-/under-prediction rates and operationally adjusted accuracy reflect the asymmetric cost of false reassurance, while dangerous-class recall measures performance on the critical minority class. We also report Clopper--Pearson intervals to quantify uncertainty around these rates. Under-prediction is the primary
safety metric.

\paragraph{Baselines.}
RF and $k$-NN were selected as stable, interpretable base learners
that expose inspectable belief states. For non-agentic comparison, we evaluate class-balanced LightGBM~\cite{ke2017lgbm}, XGBoost~\cite{chen2016xgboost}, and CatBoost~\cite{prokhorenkova2018catboost} on the identical feature set, plus cost-sensitive LightGBM variants with cost-sensitivity parameter $\alpha \in \{2,3,5,50\}$ to assess whether ARAT's architectural safeguards can be replicated by loss-function tuning.
To contextualise the routing architecture against naive aggregation,
we also evaluate soft voting (averaged probabilities) over the same RF+$k$-NN
agent pair. Code, preprocessing pipelines, and pinned result outputs are publicly available\footnote{\url{https://github.com/McDonnelletal/arat-paper}}.

\section{Results}
\label{sec:results}

\subsection{Core Accuracy and Safety Metrics}
\label{sec:core}

Table~\ref{tab:baseline} compares baselines on UNSW-NB15. Soft voting over RF+$k$-NN yields 4.80\% under-prediction. ARAT v2 reduces this to 1.70\% and raises adjusted accuracy to 98.3\%. The 1.70\% figure assumes perfect analyst resolution of escalated cases; the conditional row (93.21\%) reports automated performance only. Cost-sensitive LightGBM still underperforms ARAT on under-prediction, confirming that loss tuning does not reproduce the architectural safeguard. ARAT also improves High-class recall (Hi-R) over cost-sensitive LGB while keeping lower under-prediction. CatBoost achieves competitive under-prediction but provides no escalation queue,
correlated-failure detection, or per-layer routing audit; its Hi-R advantage
is a loss-function property, not an architectural one.
\begin{table}[H]
\centering
\small
\setlength{\tabcolsep}{3.5pt}
\renewcommand{\arraystretch}{0.92}
\caption{UNSW-NB15 baseline comparison ($n=82{,}332$, 30 features). CP$_{95}$ denotes the 95\% Clopper--Pearson interval on adjusted accuracy. Cost-sensitive LGB is a loss-tuned baseline only.}
\label{tab:baseline}
\begin{tabular*}{\linewidth}{@{\extracolsep{\fill}}lrrrrl@{}}
\toprule
System & Exact & Under & Adj & Hi-R & CP$_{95}$ (Adj) \\
\midrule
LightGBM (balanced)   & 80.6\% & 4.61\% & 95.39\% & 78.0\% & [95.2\%, 95.5\%] \\
XGBoost (balanced)    & 80.1\% & 4.97\% & 95.03\% & 85.8\% & [94.9\%, 95.2\%] \\
CatBoost (balanced)   & 79.5\% & 2.13\% & 97.87\% & 96.2\% & [97.8\%, 98.0\%] \\
RF only (tuned)       & 80.7\% & 3.76\% & 96.24\% & 72.3\% & [96.1\%, 96.4\%] \\
$k$-NN only           & 80.3\% & 6.10\% & 93.90\% & 25.6\% & [93.7\%, 94.1\%] \\
LGB cost-sensitive$^\dagger$ ($\alpha=3$) & 82.5\% & 3.35\% & 96.65\% & 69.0\% & [96.5\%, 96.8\%] \\
\midrule
\multicolumn{6}{@{}l}{\textit{Naive MAS aggregation (same RF+$k$-NN pair, no routing)}} \\
Soft vote (avg probability) & 80.8\% & 4.80\% & 95.20\% & 51.0\% & [95.1\%, 95.4\%] \\
\midrule
\multicolumn{6}{@{}l}{\textit{ARAT: 6.79\% escalated to mandatory review}} \\
ARAT v2 (all routing layers, 100\%) & 73.6\% & \textbf{1.70\%} & \textbf{98.30\%} & 74.4\% & [98.2\%, 98.4\%] \\
ARAT v2 (93.21\%$^{\ddagger}$) & 78.4\% & \textbf{1.75\%} & \textbf{98.25\%} & 74.8\% & [98.2\%, 98.3\%] \\
\bottomrule
\end{tabular*}
\begin{flushleft}
{\footnotesize $^\dagger$ Loss-tuned only; no correlated-failure detection, safety flag, or escalation queue.}\\
{\footnotesize $^{\ddagger}$ Metrics conditional on the 93.21\% of test cases not escalated to mandatory review ($n_{\text{auto}} = 76{,}742$).}
\end{flushleft}
\end{table}

\subsection{Error Dependence and Correlated Agreement Blindness}
\label{sec:dependence}

Table~\ref{tab:dependence} summarises error dependence between $IA_1$
and $AA_2$ on the holdout. The observed joint error rate is 0.134,
whereas independence would predict only 0.038, yielding a
3.53$\times$ inflation (BCa 95\% CI: [3.50, 3.59]) and strong
positive association ($\phi=0.612$, BCa 95\% CI: [0.605, 0.619]).
These intervals exclude sampling artefacts. Of all ARAT errors, 57.2\%
occur under agent agreement; the disagreement signal is blind to them.
Of all dangerous under-predictions, 90.6\% occur under agent agreement
after conservative override, confirming that disagreement-based
monitoring is structurally blind to the majority of dangerous failures.

\begin{table}[!htbp]
\vspace{4pt}
\centering
\small
\setlength{\tabcolsep}{4pt}
\renewcommand{\arraystretch}{0.95}
\caption{Error dependence between $IA_1$ and $AA_2$ on the $n=82{,}332$
holdout. BCa~=~bias-corrected and accelerated bootstrap.}
\label{tab:dependence}
\begin{tabular}{@{}lr@{}}
\toprule
\multicolumn{2}{@{}l}{\textit{Error dependence}} \\
\midrule
$P(A_1 \text{ wrong})$ & 0.193 \\
$P(A_2 \text{ wrong})$ & 0.197 \\
$P(\text{both wrong})$ & 0.134 \\
Expected joint error under independence & 0.038 \\
Observed / expected (inflation) & 3.53$\times$~[3.50, 3.59] \\
$\phi$ correlation coefficient & 0.612~[0.605, 0.619] \\
Total errors & 17,000 \\
All errors under agreement & 57.2\% (9,722/17,000) \\
Under-predictions (total) & 1,793 \\
Under-predictions under agreement (post-override) & 90.6\% (1,625/1,793) \\
\bottomrule
\end{tabular}
\end{table}

\subsection{Architectural Decomposition: Override and Safety Flag}
\label{sec:override}

The conservative override is the dominant architectural contributor, eliminating 92.8\% of disagreement-based under-predictions by construction. The unanimous-Normal safety flag (both agents predict Normal, entropy $> \theta=0.10$) contributes a more modest $-0.48$pp; its significance is mechanistic rather than aggregate: it is the only mechanism that operates when agents reach a high-entropy consensus on the `Normal' class. On this partition ($0.51\%$ High-severity prevalence), the safety flag captures $100\%$ of the dangerous under-predictions that reside within the unanimous-Normal subpopulation, proving it serves as a \textit{principled} final safety net where disagreement-based monitoring is structurally blind.

\subsection{Escalation Model and Routing Signals}
\label{sec:routing}

The escalation meta-model (logistic regression, 10 routing-time
features, isotonic calibration) achieves AUROC 0.924 and is used
solely for analyst queue ordering rather than as a hard safety gate.
Restricting evaluation to the unanimous-Normal subset ($n =
25{,}642$, a superset of the 5,590 cases flagged by the entropy
threshold) gives conditional AUROC 0.933, confirming meaningful
ranking within the critical subpopulation; the primary driver is RF
confidence (coefficient $-4.47$). Within this subset, at a 15\%
global review budget, 91.7\% of dangerous under-predictions are
retained (CP$_{95}$: [88.8\%, 94.1\%]; $k = 411$ of $n = 448$),
confirming that critical concentration sits within the highest-priority portion
of the escalation queue.

\subsection{Ablation: RF Configuration, Confusion Matrix, and Disagreement Signal}
\label{sec:ablation}
\begin{figure}[!htbp]
\vspace{-4pt}
\centering
\includegraphics[width=0.72\linewidth]{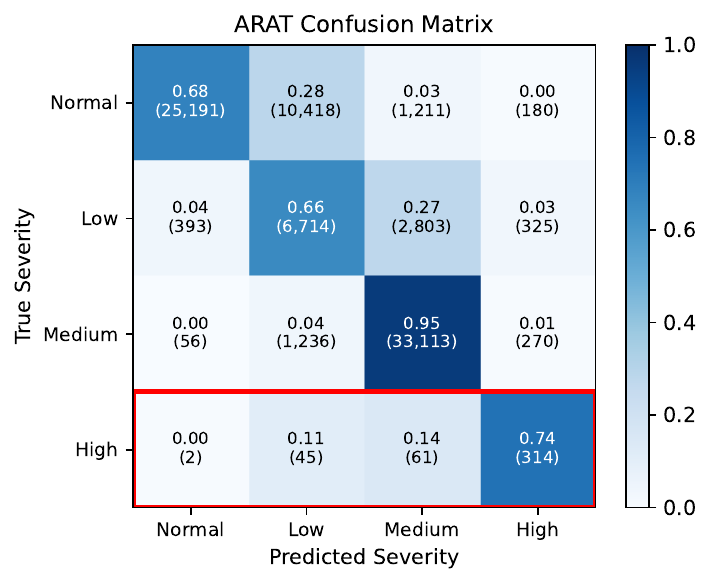}
\caption{Row-normalised confusion matrix for ARAT v2 on the UNSW-NB15 holdout set ($n = 82{,}332$). High-severity recall is 74.4\%, bounded by the agreement ceiling discussed in Section~\ref{sec:discussion}; Medium-severity recall is 95.5\%, reflecting clearer class separation.}
\label{fig:confusion}
\vspace{4pt}

\end{figure}

Figure~\ref{fig:confusion} shows the row-normalised confusion matrix for ARAT v2 on the UNSW-NB15 holdout set ($n = 82{,}332$). The red-highlighted High-severity row marks the most safety-critical class and shows 74.4\% recall, reflecting the agreement ceiling discussed in Section~\ref{sec:discussion}; Medium-severity recall of 95.5\% indicates clearer separation at that class. The Normal row's 28\% misclassification into Low reflects feature-space overlap
at the benign/low-severity boundary~\cite{moustafa2015unsw,moustafa2016evaluation}; as
over-predictions, these do not affect the under-prediction safety metric.

To probe how RF strength affects error correlation and disagreement, we compare a lighter baseline (v1) to a heavier variant (v2) on UNSW-NB15 and Diabetes, as defined in Table~\ref{tab:ablation}. On UNSW-NB15, the heavier configuration improves exact accuracy by 0.47pp, increases $\phi$ (0.584$\to$0.612), reduces disagreement (15.31\%$\to$13.75\%),
and concentrates more errors under agreement (55.46\%$\to$57.19\%):
all four mechanism indicators (error correlation $\phi$,
disagreement, errors-under-agreement, and under-prediction) move in
the predicted direction. The modest rise in under-prediction ($+$0.35pp) alongside reduced disagreement demonstrates the mechanism: a stronger, more correlated RF suppresses disagreement-triggered overrides, increasing under-prediction even as accuracy improves. On Diabetes, all four indicators move in the same direction.

\begin{table}[!htbp]
\vspace{4pt}
\centering
\small
\setlength{\tabcolsep}{4pt}
\renewcommand{\arraystretch}{0.95}
\caption{Effect of RF configuration (v1: lighter, v2: heavier) on ARAT mechanism indicators. UNSW-NB15: RF grows from 200 to 500 trees, features reduced from 37 to 30. Diabetes: RF grows from 100 to 500 trees, depth fixed at 15. Mechanism indicators move in the predicted direction on both datasets.}
\label{tab:ablation}
\begin{tabular}{@{}lcccccc@{}}
\toprule
& \multicolumn{3}{c}{UNSW-NB15} & \multicolumn{3}{c}{Diabetes} \\
\cmidrule(lr){2-4} \cmidrule(lr){5-7}
Metric & v1 & v2 & $\Delta$ & v1 & v2 & $\Delta$ \\
\midrule
RF exact accuracy   & 80.27\% & 80.74\% & $+$0.47pp & 48.32\% & 48.72\% & $+$0.40pp \\
ARAT under-prediction$^{*}$ & 1.83\% & 2.18\% & $+$0.35pp & 13.70\% & 13.83\% & $+$0.13pp \\
Dangerous-class recall & 72.51\% & 74.41\% & $+$1.90pp & 28.61\% & 28.01\% & $-$0.60pp \\
\midrule
$\phi$ error correlation & 0.584 & 0.612 & $+$0.028 & 0.279 & 0.283 & $+$0.004 \\
Disagreement rate        & 15.31\% & 13.75\% & $-$1.56pp & 45.31\% & 44.97\% & $-$0.34pp \\
Errors under agreement   & 55.46\% & 57.19\% & $+$1.73pp & 38.98\% & 39.26\% & $+$0.28pp \\
Under-preds under agreement  & 90.62\% & 90.63\% & $+$0.01pp & 79.66\% & 80.14\% & $+$0.48pp \\
\bottomrule
\end{tabular}
\begin{flushleft}
{\footnotesize $^{*}$ Routing-layer output prior to escalation; full ARAT v2 achieves
1.70\% (Table~\ref{tab:baseline}).}
\end{flushleft}
\end{table}
\vspace{-8pt}

\subsection{Cross-Dataset Validation: Diabetes Readmission Triage}
\label{sec:diabetes}

UNSW-NB15 combines real background traffic with synthetically generated attacks, whereas the UCI Diabetes 130-US Hospitals dataset is derived from real clinical records. To test whether the mechanism generalises beyond network security, and
to probe the limits of diversification as a mitigation strategy, we
replicate on the Diabetes benchmark. This is explicitly a
boundary-condition experiment, not a second performance claim; the
goal is to confirm that correlated agreement blindness is a structural
property of disagreement-based escalation, and to determine when the
architectural remedy works.

Table~\ref{tab:ablation} (Diabetes columns) reports mechanism indicators.
All four replicate: $\phi$ rises from 0.279 to 0.283 (BCa 95\% CI
v2: [0.270, 0.296]), disagreement falls by 0.34pp, errors-under-agreement
rises, and under-predictions-under-agreement rises, confirming the
mechanism operates across domains, independently of feature space,
base learner configuration, and synthetic vs.\ real data. Inflation
is 1.30$\times$ (BCa 95\% CI: [1.28, 1.31]), statistically confirmed
above independence. The escalation meta-model achieves AUROC 0.727 globally and 0.593 conditionally, above chance but weaker than UNSW-NB15, consistent with 45\% agent disagreement reducing the unanimous-Normal subpopulation. Under-prediction is essentially flat from v1 to v2 ($+$0.13pp), which is consistent
with the mechanism finding: correlated agreement blindness operates
at the level of error correlation structure, independent of whether
accuracy gains propagate to safety.

\paragraph{SVM substitution.}
Replacing $k$-NN with an SVM shifts the ARAT mechanism in opposite ways across domains. We define productive disagreement as disagreement that is associated with an error asymmetry between agents, such that the conservative override can replace a wrong prediction with a safer one. In the UNSW-NB15 dataset, substituting SVM for $k$-NN reduced the error correlation ($\phi$: $0.6117 \rightarrow 0.5180$) and increased the disagreement ($13.75\% \rightarrow 23.07\%$), reducing the under-prediction by 0.64pp and improving the adjusted accuracy from $97.82\%$ to $98.46\%$, despite a decrease in exact accuracy from $79.35\%$ to $72.92\%$. In the case of the Diabetes dataset, SVM converges with RF ($\phi$: $0.283 \rightarrow 0.696$; disagreement: $45.0\% \rightarrow 21.4\%$), reducing productive disagreement and degrading under-prediction by 3.1pp. This pattern is consistent with how the two models partition the feature space: RF and SVM appear to fail on the same noisy boundary rather than differing only in hyperparameter choice. This confirms that agent diversification is necessary but not sufficient: the safety benefit scales with whether the substituted agent produces complementary errors, not merely different inductive biases in principle.
\section{Discussion}
\label{sec:discussion}

\subsection{Recall, Agreement Ceiling, and Shared Bias}

ARAT's High-class recall of 74.4\% (CP$_{95}$: [70.0\%, 78.5\%]) is
bounded by an agreement ceiling rather than a tuning failure: 90.6\%
of dangerous under-predictions occur under agent agreement, where
routing cannot recover errors made by both base models. The missed High-class cases represent 108 samples out of 82,332 total; the 1.70\% under-prediction headline captures the
full dangerous-miss scope across all severity classes (1,793 cases
total), not fixating on the High class alone. Cost-sensitive
LightGBM's High-class recall (69.0\%) remains below ARAT's (74.4\%)
at a substantially worse under-prediction rate (3.35\% vs.\ 1.70\%),
confirming that single-model loss tuning cannot resolve errors arising
from correlated feature-space ambiguity. As only 9.4\% of under-predictions occur under disagreement, routing can theoretically recover at most that fraction; 74.4\% Hi-R indicates near-complete recovery within the architecturally recoverable subset.

\subsection{Broader Implications for Multi-Agent Safety}
These cross-dataset results, spanning synthetic-plus-real traffic and real clinical data,
suggest a structural rather than dataset-specific effect.
Correlated agreement blindness is not unique to these domains. Inter-agent
misalignment bias is a systematic failure mode in multi-agent LLM
systems~\cite{cemri2025multi}; correlated errors intensify as capability
increases~\cite{goel2025great}; monoculture collapse is a recognised
vulnerability in governed MAS~\cite{reid2025risk}. In any MAS where a
disagreement signal gates human oversight, the signal's reliability
degrades as agents become more capable and more correlated. This
applies directly to agentic LLM pipelines: as foundation models converge
on shared pretraining and fine-tuning, the inductive bias diversity that
disagreement-based escalation depends on erodes silently. The SVM
substitution experiment operationalises this: diversification by
inductive bias alone is insufficient if the new agent fails at the same
feature-space boundary. A pre-deployment diagnostic (measuring $\phi$
and productive-disagreement rate across candidate agent pairs before
committing to an architecture) would provide a principled basis for
agent selection in safety-critical settings.

\section{Limitations}
\label{sec:limitations}
Three limitations bound our claims. First, the headline 1.70\%
under-prediction rate credits the analyst with correctly resolving the
escalated 6.79\% of cases; the conditional row of
Table~\ref{tab:baseline} (1.75\% on the 93.21\% decided automatically)
bounds the automated system alone, and realised performance lies
between the two. Second, ARAT is demonstrated with two deliberately
simple, interpretable base learners on two tabular datasets; the SVM
substitution shows the override's benefit depends on the agent pair
sustaining productive disagreement and can vanish under convergence,
so we offer $\phi$ together with the productive-disagreement rate as a
pre-deployment diagnostic, not an unconditional guarantee.
Finally, we benchmark against single-model, loss-tuned, and naive
aggregation baselines but not learned multi-agent coordination
frameworks, which ARAT deliberately forgoes in favour of auditable
fixed routing; that comparison is future work.

\section{Conclusion}
\label{sec:conclusion}

ARAT achieves 98.3\% operationally adjusted accuracy and 1.70\%
dangerous under-prediction on 82,332 UNSW-NB15 holdout samples.
A per-layer decomposition confirms that the conservative override
contributes $-$2.6pp and the safety-flag gate a further $-$0.5pp
over naive soft voting, gains not reproduced by loss-function tuning alone.
The primary contribution is the characterisation of correlated
agreement blindness as a structural property of disagreement-triggered
escalation: under correlated failure, escalation becomes blind to the
failures that most require human review, and this blindness intensifies
as base learner strength increases. Cross-dataset replication supports these
indicators; SVM substitution further shows that agent diversification
improves safety only when it generates productive disagreement,
providing practitioners with a concrete pre-deployment diagnostic.
As agentic pipelines deploy increasingly capable and correlated models,
disagreement-triggered monitoring will become less reliable precisely
where it is most needed. The phenomenon we term \textit{correlated agreement blindness} necessitates explicit
architectural consideration: not as an edge case, but rather as an expected
consequence of delivering the agent improvements that safety monitoring depends upon.

\newpage
\section*{Acknowledgements}

This research was conducted with the financial support of Research
Ireland at ADAPT, the Research Ireland Centre for AI-Driven Digital
Content Technology at Trinity College Dublin, 13/RC/2106\_P2. For the purpose of Open
Access, the author has applied a CC BY public copyright licence to any
Author Accepted Manuscript version arising from this submission.

This work has been funded by CKDelta through the establishment of an
AI research collaboration between Trinity College Dublin and CKDelta.

\end{document}